\documentclass[twoside,twocolumn,9pt]{article}
\usepackage{extsizes}
\usepackage[super,sort&compress,comma]{natbib}
\usepackage[version=3]{mhchem}
\usepackage[left=1.5cm, right=1.5cm, top=1.785cm, bottom=2.0cm]{geometry}
\usepackage{balance}
\usepackage{mathptmx}
\usepackage{amssymb}
\usepackage{sectsty}
\usepackage{graphicx}
\usepackage{lastpage}
\usepackage[format=plain,justification=justified,singlelinecheck=false,font={stretch=1.125,small,sf},labelfont=bf,labelsep=space]{caption}
\usepackage{float}
\usepackage{fancyhdr}
\usepackage{fnpos}
\usepackage[english]{babel}
\addto{\captionsenglish}{%
    
}
\usepackage{array}
\usepackage{droidsans}
\usepackage{charter}
\usepackage[T1]{fontenc}
\usepackage[usenames,dvipsnames]{xcolor}
\usepackage{setspace}
\usepackage[compact]{titlesec}
\usepackage[hidelinks]{hyperref}
\definecolor{cream}{RGB}{222,217,201}

\newcommand{\N}{\bar{n}}

\begin{document}

    \pagestyle{fancy}
    \thispagestyle{plain}
    \fancypagestyle{plain}{
        %%%HEADER%%%
        \renewcommand{\headrulewidth}{0pt}
    }
    %%%END OF HEADER%%%

    %%%PAGE SETUP - Please do not change any commands within this section%%%
    \makeFNbottom
    \makeatletter
    \renewcommand\LARGE{\@setfontsize\LARGE{15pt}{17}}
    \renewcommand\Large{\@setfontsize\Large{12pt}{14}}
    \renewcommand\large{\@setfontsize\large{10pt}{12}}
    \renewcommand\footnotesize{\@setfontsize\footnotesize{7pt}{10}}
    \makeatother

    \renewcommand{\thefootnote}{\fnsymbol{footnote}}
    \renewcommand\footnoterule{\vspace*{1pt}%
        \color{cream}\hrule width 3.5in height 0.4pt \color{black}\vspace*{5pt}}
    \setcounter{secnumdepth}{5}

    \makeatletter
    \renewcommand\@biblabel[1]{#1}
    \renewcommand\@makefntext[1]%
    {\noindent\makebox[0pt][r]{\@thefnmark\,}#1}
    \makeatother
    \renewcommand{\figurename}{\small{Fig.}~}
    \sectionfont{\sffamily\Large}
    \subsectionfont{\normalsize}
    \subsubsectionfont{\bf}
    \setstretch{1.125} %In particular, please do not alter this line.
    \setlength{\skip\footins}{0.8cm}
    \setlength{\footnotesep}{0.25cm}
    \setlength{\jot}{10pt}
    \titlespacing*{\section}{0pt}{4pt}{4pt}
    \titlespacing*{\subsection}{0pt}{15pt}{1pt}
    %%%END OF PAGE SETUP%%%

    %%%FOOTER%%%
    \fancyfoot{}
    %%ArXiv%%\fancyfoot[LO,RE]{\vspace{-7.1pt}\includegraphics[height=9pt]{head_foot/LF}}
    %%ArXiv%%\fancyfoot[CO]{\vspace{-7.1pt}\hspace{13.2cm}\includegraphics{head_foot/RF}}
    %%ArXiv%%\fancyfoot[CE]{\vspace{-7.2pt}\hspace{-14.2cm}\includegraphics{head_foot/RF}}
    \fancyfoot[RO]{\footnotesize{\sffamily{1--\pageref{LastPage} ~\textbar  \hspace{2pt}\thepage}}}
    \fancyfoot[LE]{\footnotesize{\sffamily{\thepage~\textbar\hspace{3.45cm} 1--\pageref{LastPage}}}}
    \fancyhead{}
    \renewcommand{\headrulewidth}{0pt}
    \renewcommand{\footrulewidth}{0pt}
    \setlength{\arrayrulewidth}{1pt}
    \setlength{\columnsep}{6.5mm}
    \setlength\bibsep{1pt}
    %%%END OF FOOTER%%%

    %%%FIGURE SETUP - please do not change any commands within this section%%%
    \makeatletter
    \newlength{\figrulesep}
    \setlength{\figrulesep}{0.5\textfloatsep}

    \newcommand{\topfigrule}{\vspace*{-1pt}%
        \noindent{\color{cream}\rule[-\figrulesep]{\columnwidth}{1.5pt}} }

    \newcommand{\botfigrule}{\vspace*{-2pt}%
        \noindent{\color{cream}\rule[\figrulesep]{\columnwidth}{1.5pt}} }

    \newcommand{\dblfigrule}{\vspace*{-1pt}%
        \noindent{\color{cream}\rule[-\figrulesep]{\textwidth}{1.5pt}} }

    \makeatother
    %%%END OF FIGURE SETUP%%%

    %%%TITLE, AUTHORS AND ABSTRACT%%%
    \twocolumn[
\begin{@twocolumnfalse}
    %%ArXiv%%{\includegraphics[height=30pt]{head_foot/SM}\hfill\raisebox{0pt}[0pt][0pt]{\includegraphics[height=55pt]{head_foot/RSC_LOGO_CMYK}}\\[1ex]
        %%ArXiv%%    \includegraphics[width=18.5cm]{head_foot/header_bar}}
    \par
    \vspace{1em}
    \sffamily
    \begin{tabular}{m{4.5cm} p{13.5cm} }

        %%ArXiv%%\includegraphics{head_foot/DOI}
        & \noindent\LARGE{\textbf{Effect of Photon Counting Shot Noise on Total Internal Reflection Microscopy}} \\
        \vspace{0.3cm} & \vspace{0.3cm} \\

        & \noindent\large{Fan Cui\textit{$^{a}$} and David J. Pine\textit{$^{ab}$}} \\%Author names go here instead of "Full name", etc.

        %%ArXiv%%\includegraphics{head_foot/dates}
            & \noindent\normalsize{
                Total internal reflection microscopy (TIRM) measures changes in the distance between a colloidal particle and a transparent substrate by measuring the scattering intensity of the particle illuminated by an evanescent wave.
                From the distribution of the recorded separation distances, the height-dependent effective potential $\varphi(z)$ between the colloidal particle and the substrate can be measured.
                In this work, we show that spatial resolution with which TIRM can measure $\varphi(z)$ is limited by the photon counting statistics of the scattered laser light.
                We develop a model to evaluate the effect of photon counting statistics on different potential profiles using Brownian Dynamics simulations and experiments.
                Our results show that the effect of photon counting statistics depends on spatial gradients $\partial \varphi/\partial z$ of the potential, with the result that sharp features tend to be significantly blurred.
                We further establish the critical role of photon counting statistics and the intensity integration time $\tau$ in TIRM measurements, which is a trade-off between narrowing the width of the photon counting distribution and capturing the instantaneous position of the probe particle.} \\
        \end{tabular}

    \end{@twocolumnfalse} \vspace{0.6cm}

    ]
    %%%END OF TITLE, AUTHORS AND ABSTRACT%%%

    %%%FONT SETUP - please do not change any commands within this section
    \renewcommand*\rmdefault{bch}\normalfont\upshape
    \rmfamily
    \section*{}
    \vspace{-1cm}

    %%%FOOTNOTES%%%

    \footnotetext{\textit{$^{a}$~Department of Physics, Center for Soft Matter Research, New York University, 726 Broadway, New York, NY 10003, USA. E-mail: cuifan@nyu.edu, pine@nyu.edu}}
    \footnotetext{\textit{$^{b}$~Department of Chemical \& Biomolecular Engineering, Tandon School of Engineering,  New York University, 6 MetroTech Center, Brooklyn, New York 11201, USA.}}
    \footnotetext{\dag~Electronic Supplementary Information (ESI) available: [details of any supplementary information available should be included here]. See DOI: 10.1039/cXsm00000x/}

    %%%END OF FOOTNOTES%%%

    %%%MAIN TEXT%%%%
    \section{\label{sec:intro}Introduction}

    Total internal reflection microscopy\cite{prieve1999tirm} (TIRM) is a powerful method for measuring the microscopic interactions of colloidal particles in a liquid suspension.
    Since its development some 30 years ago,\cite{prieve1990tirm} it has been used to measure various colloidal interactions, including screened electrostatic repulsion,\cite{prieve1999tirm,bike2000tirm} steric repulsion due to grafted or adsorbed polymers,\cite{prieve2000tirmbevan,ngai2014tirm,prlang2007steric} van der Waals attraction,\cite{prieve2000tirmbevan} depletion attraction,\cite{bevan2012tirmdepletion,depletionprlang} critical Casimir interactions\cite{bechinger2008tirmcasimir}, and interactions of DNA-coated colloids.\cite{rogers2021dnaccMembrane}
    Spatial resolutions as small as 1~nm have been reported.\cite{prieve1990tirm}
    As such, TIRM has become an invaluable tool for understanding colloidal interactions at a microscopic scale.

    In spite of TIRM's long and enduring use, the effects of photon counting statistics, often called \textit{shot noise}, on TIRM measurements of colloidal interaction potentials have not been fully worked out.
    While shot noise has been considered for the special case of a particle is confined by optical tweezers in an evanescent field\cite{clapp1999noise,clapp2001noise} and for microrheology measurements in an evanescent field,\cite{helseth2004fundamental} in conventional TIRM measurements, shot noise is generally regarded as insignificant without further detailed consideration.\cite{frej1993hindered,waltzmentionshot2005new}
    Indeed, shot noise can often be ignored when measuring potentials with soft features like a double-layer potential at low ionic strength.\cite{frej1993hindered,waltzmentionshot2005new}
    However, as we show in this paper, shot noise can be the limiting factor when measuring interactions with hard core or short-range potential profiles and for particles with fast dynamics.
    In this work, we systematically study the effects of shot noise on TIRM measurements of interaction potentials and identify the potential profile features that are most prone to corruption by shot noise.
    We also provide the means to quantitatively determine and minimize how it distorts the measurement of the potential.

    \begin{figure}[b!]
        \centering
        \includegraphics{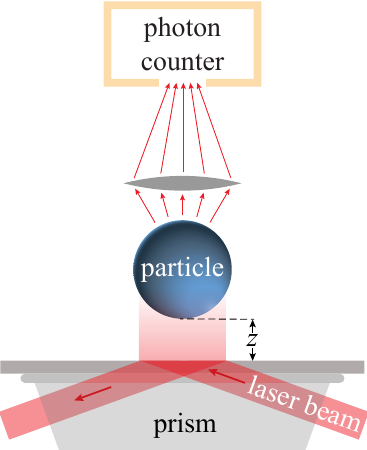}
        \caption{TIRM schematic.
            Laser light enters from the right and is totally internally reflected at the substrate, launching an exponentially-damped evanescent wave towards the particle, which is a distance $z$ above the substrate in a liquid suspension.
            Light scattered by the particle is collected by a microscope objective lens and directed toward a photon counter.}
        \label{fig:tirmschematic}
    \end{figure}

    Fig.\ \ref{fig:tirmschematic} shows a schematic of a typical TIRM experiment and summarizes the basic experimental setup.
    The TIRM technique is based on two simple ideas.
    The first idea is that the probability that a particle at equilibrium in a liquid suspension is at a height $z$ above the substrate is given by the Boltzmann distribution
    \begin{align}
        p(z) = A\,e^{-\varphi(z)/k_BT} \;,
        \label{eq:boltzmann}
    \end{align}
    where $\varphi(z)$ is the effective interaction potential between the particle and the substrate.
    Inverting this equation, we obtain the expression
    \begin{align}
        \frac{\varphi(z) -\varphi(z_r)}{k_BT} = \ln \frac{p(z_r)}{p(z)} \;,
        \label{eq:boltzmannInv}
    \end{align}
    where we have introduced a reference height $z_r$ to eliminate the normalization factor $A$ that appears in eqn \eqref{eq:boltzmann}.
    Equation \eqref{eq:boltzmannInv} tells us that if we can measure the distribution of particle heights $p(z)$ above the substrate, we can determine the effective interaction between the particle and the substrate.

    The second idea starts with the observation that the evanescent intensity $I_e$ of light totally internally reflected from the substrate decays exponentially with height above the substrate, $I_e(z) = I_{e0}\,\exp(-\beta z)$,
    where $\beta^{-1}$ is the penetration depth of the evanescent wave, which is typically in the range of 70-200 nm.
    Because the intensity $I$ of the light scattered by the particle is proportional to the intensity $I_e$ of the evanescent field that is incident on the particle, the scattered intensity also depends exponentially on the height of the particle,\cite{kerker79evanescent,prieve1999tirm}
    \begin{align}
        I(z) = I_0\,e^{-\beta z} \;.
        \label{eq:scatint}
    \end{align}
    Thus, we see that the scattered intensity is related directly to the height $z$ of a particle.
    This gives us a way to determine the height $z$ of the particle.
    We note that the exponential form for the intensity given by Eq.\ \ref{eq:scatint} is stricly correct only if certain precautions are taken in the experimental design.\cite{helden2006single}
    In what follows, we assume such precautions have been taken.

    The probability $P(I)\,dI$ that the scattered intensity is between $I$ and $I+dI$ is the equal to the probability $p(z)\,dz$ that the particle is between a height of $z$ and $z+dz$, and thus are related by
    \begin{align}
        P(I)\,|dI| = p(z)\,|dz| \;.
        \label{eq:probIz}
    \end{align}
    Solving for $p(z)$ yields,
    \begin{align}
        p(z) = P(I) \left|\frac{dI}{dz}\right| = \beta\, P(I)\, I(z) \;,
        \label{eq:pz}
    \end{align}
    where we have used eqn \eqref{eq:scatint} to evaluate the derivative.
    Thus, the probability distribution of heights $p(z)$ appearing in eqn \eqref{eq:boltzmannInv} can be expressed in terms of the probability distribution of scattered intensities $P(I)$.

\begin{figure}[!ht]
        \includegraphics{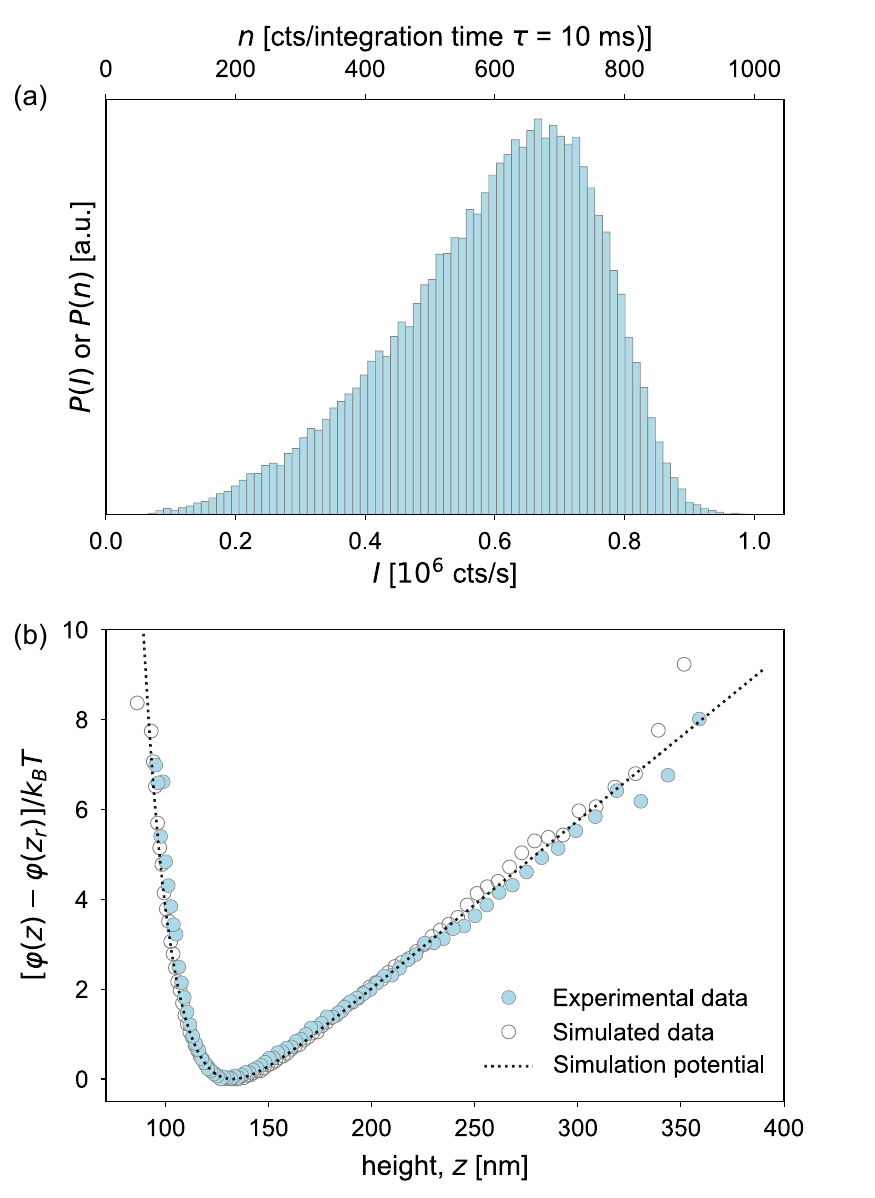}
        \caption{
            \label{fig:Sim_Demo}
            TIRM measurement of the interaction potential for a negatively charged polystyrene sphere with diameter of 8~$\mu$m in 0.5 mM NaCl aqueous solution above a negatively charged glass substrate.
            (a) Histogram of scattered intensities obtained from a TIRM experiment with an integration time $\tau = 10$~ms.
            (b) Solid circles show the experimentally measured interaction potential $\varphi(z)$, with the potential minimum height aligned at $z_r$ = 133 nm.
            A least-square-fit to the data gives a Debye length of $\kappa^{-1} = 12.1$~nm and  $G = 0.146$~pN.
            Open circles show the potential profile obtained from a Brownian dynamics simulation using $\Delta t = 0.2$~ms as simulation step size with 12.5 million steps and integration time $\tau = 10$~ms.
            A least squares fit gives $\kappa^{-1} = 13.5$~nm and $G = 0.157$~pN.
            Dotted line shows potential $\varphi(z)$ used as an input for the simulation with $\kappa^{-1} = 13.7$~nm and $G = 0.152$~pN.
        }
    \end{figure}

    In a TIRM experiment, changes in the scattered intensity are monitored, typically for ten minutes or more, by repeatedly counting photons over some short interval of time $\tau$, the \textit{integration time}, typically on the order of milliseconds.
    From this chain of measurements, a histogram of scattered intensities $N(I)$ is constructed, where $N(I)$ is the number of observations of intensity between $I$ and $I+\Delta I$.
    For a sufficiently large number of measurements $N(I) \propto P(I)$.
    Fig.\ \ref{fig:Sim_Demo}a shows such a histogram obtained from a TIRM measurement of a negatively charged polystyrene sphere in aqueous suspension above a negatively charged glass substrate.

    Using eqn \eqref{eq:pz} for $p(z)$, eqn \eqref{eq:boltzmannInv} can be rewritten as
    \begin{align}
        \frac{\varphi(z) -\varphi(z_r)}{k_BT} = \ln \frac{N(I_r)\,I(z_r)}{N(I)\,I(z)} \;,
        \label{eq:NTIRM}
    \end{align}
    where $P(I)$ has been replaced by $N(I)$, which is valid if $\Delta I$ is small and the number of of samples $N$ is large.
    Fig.\ \ref{fig:Sim_Demo}(b) shows the potential $\varphi(z)$ obtained from the histogram of scattered intensities shown in Fig.\ \ref{fig:Sim_Demo}a.

    \section{Shot noise}\label{sec:shotnoise}

    Like most optical measurements, the resolution and accuracy of TIRM are limited by instrumental noise.
    The types of noise commonly identified are background scattering, laser power fluctuations, statistical noise, and photon detection shot noise.\cite{frej1993hindered,prieve_remove_noise,prieve2000shollSimulation,clapp2001noise,noisevdwnayeri2013,noisesoftenstrongforce2013Bevan}
    In earlier efforts, several methods were developed to reduce the effect of background noise,\cite{prieve_remove_noise,clapp2001noise} including subtraction of averaged background intensity and applying low-pass filter of the measured signals.
    The effect of laser power fluctuations can usually be kept small \cite{prieve2000shollSimulation} in and can be managed with improved laser design.\cite{helseth2004fundamental}
    In addition, there is the statistical noise associated with forming the histogram of intensities from a finite number of measurements.\cite{prieve2000shollSimulation}
    This is the source of the deviations from the theoretical predictions at large $z$ that are visible in Fig.\ \ref{fig:Sim_Demo}.
    This source of noise can usually be made negligible by acquiring data for a sufficiently long time period or performing the same measurements for multiple times.\cite{prieve1990tirm,prieve2000shollSimulation}

    %Any set of experimental measurements is subject to noise.
    %Prieve showed that above certain level, around 5\%, both background scattering and fluctuations in the incident laser intensity could have noticeable adverse effects on the measured potential profiles.\cite{prieve2000shollSimulation,prieve_remove_noise}
    %Fortunately, with high-quality laser sources and careful optical design, laser fluctuations and background scattering can be kept well below 1\%.
    %In addition to these sources of noise, there is the statistical noise associated with forming the histogram of intensities from a finite number of measurements.
    %This is the source of the deviations from the theoretical predictions at large $z$ that are visible in Fig.\ \ref{fig:Sim_Demo}.
    %This source of noise can usually be made negligible simply by acquiring data for a sufficiently long period of time.\cite{prieve1990tirm}
    %Indeed, shot noise are inherent in any optical detection system and cannot be easily filtered out as it strongly correlate to the measurement signal.

    %\textcolor{red}{One of the lesser studied noise sources is the photon counting shot noise.
    %Shot noise is inherent in any optical detection system and cannot be filtered out as it strongly correlates to the real measurement signal.}
    The implicit assumption in the analysis of TIRM data using eqn \eqref{eq:NTIRM} is that there is a one-to-one correspondence between the scattered intensity and the particle position, which is given by eqn \eqref{eq:scatint}.
    However, for any measurement of light intensity, there are intrinsic quantum fluctuations (shot noise) associated with photon counting.
    In a typical TIRM experiment, the intensity of the scattered light is measured by some quantum mechanical photon counting process, for example, using a photomultiplier or an avalanche photodiode.
    Indeed, these are the most sensitive and information-rich methods of detecting the scattered intensity.
    In a typical TIRM experiment, the intensity is measured by counting photons for some integration time interval $\tau$, typically on the order of milliseconds.
    In this case, the probability of detecting $n$ photons in a time interval $\tau$ is given by a Poisson distribution\cite{mandel1959poissonlight}
    \begin{align}
        P_{\tau}(n;\N) = \frac{\N^n}{n!} e^{-\N} \;,
        \label{eq:photondist}
    \end{align}
    where $\N$ is the average number of photons detected in a time $\tau$ for a given constant intensity $I$.
    The width of the distribution, as measured by the square root of the variance, is $\sqrt{\N}$.
    The finite width of $P_{\tau}(n;\N)$ means that there is an intrinsic uncertainty, sometimes called \textit{shot noise}, in any measurement of the scattered intensity.
    This means that there will be an intrinsic uncertainty in the particle height and in the determination of the effective potential $\varphi(z)$.
    This limits the resolution with which TIRM can measure $\varphi(z)$.
    As we will show, for potentials that do not vary too rapidly in space, this does not pose a serious limitation.
    However, for rapidly-varying potentials, such as those exhibited by particles with a fairly hard-core repulsion or a very short-range attraction, it can pose a significant limitation.

    The intensity $I$ that appears in eqns \eqref{eq:scatint}--\eqref{eq:NTIRM} is the classical intensity, without shot noise.
    The units of intensity are arbitrary in this context, so without loss of generality we can write $n_c = I\,\tau$ and $P(I)\,dI=P(n_c)\,dn_c$, where $n_c$ is taken to mean the (classical) intensity, measured in counts per integration time, that would be measured if there were no shot noise.
    Thus, we can rewrite eqn \eqref{eq:scatint} as
    \begin{align}
        n_c(z) = n_{c0} e^{-\beta z} \;.
        \label{eq:n_c}
    \end{align}

    However, for a given integration time $\tau$ and classical intensity $n_c = I\,\tau$, the number of photons $n$ actually measured is Poisson distributed around $n_c$ according to eqn \eqref{eq:photondist} with $\N=n_c$.
    This means that the two distribution functions $P(n)$, which is measured in a standard TIRM experiment, and $P(n_c)$, which is what should be used in eqn \eqref{eq:pz}, are different.
    In the limit of a very large number of measurements, two are related by
    \begin{align}
        P(n) = \sum_{n_c} P(n_c)\, P_{\tau}(n; n_c)
             = \sum_{n_c} P(n_c)\, \frac{n_c^{n}}{n!} e^{-n_c} \;.
        \label{eq:Pcon}
    \end{align}
    As $P_{\tau}(n; n_c)$ is peaked around $n_c$ with a width $\sqrt{n_c}$, we see that the measured intensity distribution $P(n)$ is similar to (but distinct from) a discrete convolution of the classical intensity distribution $P(\bar{n})$ with the Poisson distribution $P_{\tau}(n;\N)$ given by eqn \eqref{eq:photondist}.
    Thus, any abrupt change in $P(n_c)$, which occurs when there is an abrupt change in $\varphi(z)$, will be rounded by $P_{\tau}(n; n_c)$ on a scale given by $\sqrt{n_c}$.
    This will lead to a blurring in the potential $\varphi(z)$ measured by TIRM.

    The blurring of $P(n)$ relative to $P(n_c)$ limits the resolution with which TIRM can measure a particle's height $z$.
    For a particle at height $z$, the average number of photons counted in a time $\tau$ is given by eqn \eqref{eq:n_c}.
    Taking the differential of eqn \eqref{eq:n_c}, we obtain
    \begin{align}
        \frac{dn_c}{n_c} = -\beta\, dz \;.
    \end{align}
    Setting $dn_c$ equal standard deviation of the photon counting fluctuations $\pm\sqrt{n_c}$, we obtain an expression for the uncertainty in the measured particle height due to photon shot noise
    \begin{align}
        \Delta z_m \equiv \pm\frac{\beta^{-1}}{\sqrt{n_c}} \;.
        \label{eq:error}
    \end{align}
    According to eqn \eqref{eq:error}, the estimated error $\Delta z_m$ in the measured height is the penetration depth $\beta^{-1}$ of the evanescent wave divided by the square root of the average number of photons counted during the integration time $\tau$.
    It should be noted that eqn \eqref{eq:error} provides a lower limit on the spatial resolution of a particle's position that can be inferred from a measurement of the scattered intensity.
    In a typical experiment, the maximum photon counting rate is about $10^6$~cts/s and the integration time is typically about 1~ms so that the average number of photons counted $n_c$ is about 1000.
    A typical optical penetration depth $\beta^{-1}$ is about 100~nm.
    In this case, eqn \eqref{eq:error} gives $\Delta z_m \sim 3$~nm.

    The fundamental problem with the conventional TIRM analysis is that eqn \eqref{eq:probIz} is not strictly correct.
    Because of the quantum fluctuations associated with photon counting, there is not a strict one-to-one correspondence between a measurement of the scattered intensity and the particle height $z$.
    Thus $I(z)$ in eqn \eqref{eq:NTIRM} is not a perfect proxy for particle position as assumed in eqn \eqref{eq:scatint}.
    Similarly, the $N(I)$ used in eqn \eqref{eq:NTIRM} is not a perfect proxy for $p(z)$, as assumed in eqn \eqref{eq:probIz}. In what follows, we explore the consequences of this problem through simulation and experiment and develop strategies for minimizing and mitigating the deleterious effects of photon shot noise.

    \section{\label{sec:level2}Results}

    \subsection{Brownian dynamics simulations}\label{sec:sim}

    To evaluate the effects of shot noise on the measured potential energy profiles, we first numerically simulate the trajectories of a colloidal particle and then use them to construct the noise-corrupted scattering intensities.
    We use the Brownian dynamics simulation method first described by Sholl and Prieve.\cite{prieve2000shollSimulation}
    A colloid's Brownian motion in a force field along the vertical direction can be described using a Langevin equation:
    \begin{align}
        m \frac{dv(z)}{dt} = -\zeta v(z) + \delta f(t) + F(z) \;,
        \label{eq:Langevin}
    \end{align}
    where $m$ is the mass of the particle, $v(z)$ is the velocity of the particle along the vertical direction, $\zeta$ is the friction coefficient, and $F$ is the force applied on the particle: $F = -d\varphi(z)/dz$.
    The random fluctuating force $\delta f(t)$ accounts for the interactions of the particle with the fluid in which it is suspended.
    This fluctuation has the usual zero mean and delta function correlation consistent with the fluctuation-dissipation theorem: $\langle\delta f(t)\rangle=0$, and $\langle\delta f(t)\delta f(t')\rangle=2\zeta k_BT\delta (t-t')$.

    From the Langevin equation, Ermak and McCammon\cite{displacementequation} developed a method for simulating the diffusive behavior of Brownian particles in a solution, with a displacement equation given by:
    \begin{align}
        z (t+\Delta t) = z(t) + \frac{dD}{dz}\Delta t + \frac{D}{k_BT}F(z)\Delta t + Z(\Delta t) \;,
        \label{eq:Displacementequation}
    \end{align}
    where $D$ is the particle's diffusion coefficient, and $Z(\Delta t)$ is a Gaussian random displacement with $\langle Z\rangle = 0$ and $\langle Z^2\rangle = 2D\Delta t$.

    When a particle is close to a surface, as is the case in a typical TIRM measurement, the mobility of the particle is hindered compared to its motion in a free-solution, and depends strongly on the separation distance between the particle and the surface.
    When the separation distance is small (comparable to or smaller than the particle radius $r$), the close-wall effect can be taken into account using $D(z)=\lambda D_0$, where $D_0$ is the free diffusion coefficient, and $\lambda$ is a function of $\gamma=z/r$, where $z$ is the distance between the surfaces of the substrate and the sphere.
    The function $\lambda(\gamma)$ was calculated by Brenner and is given in the form of a slowly converging infinite series.\cite{brenner_nearwalldiffusion_verticle}
    The function is well-approximated by a simplified form using a regression of the infinite-series results\cite{Prieve_hinderedDiffusion_2000}:
    \begin{align}
        \lambda = \frac{6\gamma^2+2\gamma}{6\gamma^2+9\gamma+2} \;.
        \label{eq:simplied-nearwall_D}
    \end{align}
    Using this expression and eqn \eqref{eq:Displacementequation}, we can simulate the trajectory of a colloid close to a glass wall for any known force $F(z)$.

    \subsection{Simulation of double-layer repulsion and gravity}

    We start by simulating a charge-stabilized polystyrene (PS) colloidal particle in dynamic equilibrium close to a glass surface in an ionic solution, which corresponds to the experiment we introduced in \S\ref{sec:shotnoise}.
    We assume there are only two dominant interactions: electrostatic repulsion and gravity.
    We disregard other close-range interactions such as van der Waals forces, which is a valid assumption for systems with low ionic strength and highly-charged surfaces.\cite{prieve1999tirm}
    The screened electrostatic interaction is modeled by the DLVO theory  using the Derjaguin approximation, which leads to a Yukawa potential.\cite{verwey_doublelayertheory,prieve1990tirm,Bechinger_doublelayer}
    The total potential $\varphi(z)$ is the sum of the screened electrostatic interaction and gravity:
    \begin{align}
        \varphi(z) = B e^{-\kappa z} + Gz \;,
        \label{eq:G-DLpotential}
    \end{align}
    where $G$ is the net buoyant weight of the particle, $\kappa^{-1}$ is the Debye length, and
    \begin{align*}
        B=\frac{16r}{k_BT\lambda_B} \prod_{\sigma_i = \sigma_C, \sigma_G}\tanh\left[\frac{1}{2}\sinh^{-1}{\left(\frac{2\pi\lambda_B\sigma_i}{e\kappa}\right)}\right] \;.
    \end{align*}
    Here, $\lambda_B = e^2/4\pi\epsilon k_BT$ is the Bjerrum length, $\epsilon$ is the permittivity of the solvent, and $\sigma_C$ and $\sigma_G$ are the surface charge densities of the colloid and the glass, respectively.

    In TIRM measurements, a reference potential height $z_r$ is introduced to eliminate $B$:
    \begin{align}
        \frac{\varphi(z)-\varphi(z_r)}{k_BT}=\frac{G}{k_BT\kappa}\left[e^{-\kappa(z-z_r)}-1\right]+\frac{G}{k_BT}(z-z_r) \;.
        \label{eq:tirmpot}
    \end{align}
    In our experiments, $\varphi(z_r)$ and its corresponding reference position $z_r$ are usually set to 0, which is the usual practice with TIRM experiments as the absolute distance to the substrate and parameters $B$ are typically unknown.
    In the simulations, however, it is helpful to use a reasonable value of $B$ to estimate the actual position of the colloid. In a 0.5 mM NaCl solution, $B$ is calculated to be $8.4\!\times\!10^3 k_BT$ for a 8 $\mu$m-diameter PS sphere based on literature values. \cite{prieve1999tirm}

    We use eqn \eqref{eq:Displacementequation} to generate the height trajectories $z_i$ of an 8-$\mu$m PS particle in a 0.5 mM NaCl solution, using a temporal step size of $\Delta t = 0.2$ ms (small enough to capture particle movement, see Fig.\ S2 in SI) and a total of 12.5 million steps (corresponding to a physical run time of about 40 minutes).
    The particle starting position is taken to be the distance where the potential reaches its theoretical minimum, $z_m = 133$~nm. The simulated height trajectories are shown in Fig.\ S1.

    The simulated vertical trajectories are used to generate light scattering intensity data in multiple steps.
    First, the mean number of classical counts $n_{c_i}$, uncorrupted by shot noise, is generated using eqn \eqref{eq:scatint} for each time step in the simulation
    \begin{align}
        n_c = \Delta t \, I_0 \, e^{-\beta z_i} \;,
    \end{align}
    where $I_0$ is set such that the maximum intensity corresponds to the typical maximum experimental value of about $10^6$~cts/s, which occurs when $z$ is at its point of closest approach to the substrate.
    Next, the mean number of counts $n_{c_j}$ accumulated over each integration time interval $\tau$ is generated by summing over successive integer number $k$ time steps where $\tau = k\,\Delta t$:
    \begin{align}
        n_{c_j} = \sum_{i=k(j-1)+1}^{kj} n_{c_i} \;.
    \end{align}
    These are the data that are used to construct the $P(n_c)$ histogram of intensities without shot noise.
    Finally, the sequence of intensity data $n_j$ with shot noise is constructed by selecting a random number $n_j$ from a Poisson distribution for each $n_{c_j}$:
    \begin{align}
        n_j = \mathrm{Pois}(n_{c_j}) \;.
    \end{align}
    These are the data that are used to construct the $P(n)$ histogram of intensities with shot noise.

    As shown in Fig. \ref{fig:Sim_Demo}(b), the data for $\varphi(z)$ obtained from the simulated $P(n)$ agrees well with both the experimental measurement and the analytical calculations from eqn \eqref{eq:G-DLpotential}, confirming that our simulation can quantitatively describe experimental data from TIRM measurements.
    In fact, for the parameters used here, the potential $\varphi(z)$ obtained from the simulated $P(n)$ is statistically indistinguishable from that obtained from $P(n_c)$.
    Under such circumstances, shot noise poses no problem for extracting the potenital $\varphi(z)$ using TIRM.
    As we shall see, however, this is not always the case.

    \begin{figure*}[h!]
        \centering
        \includegraphics{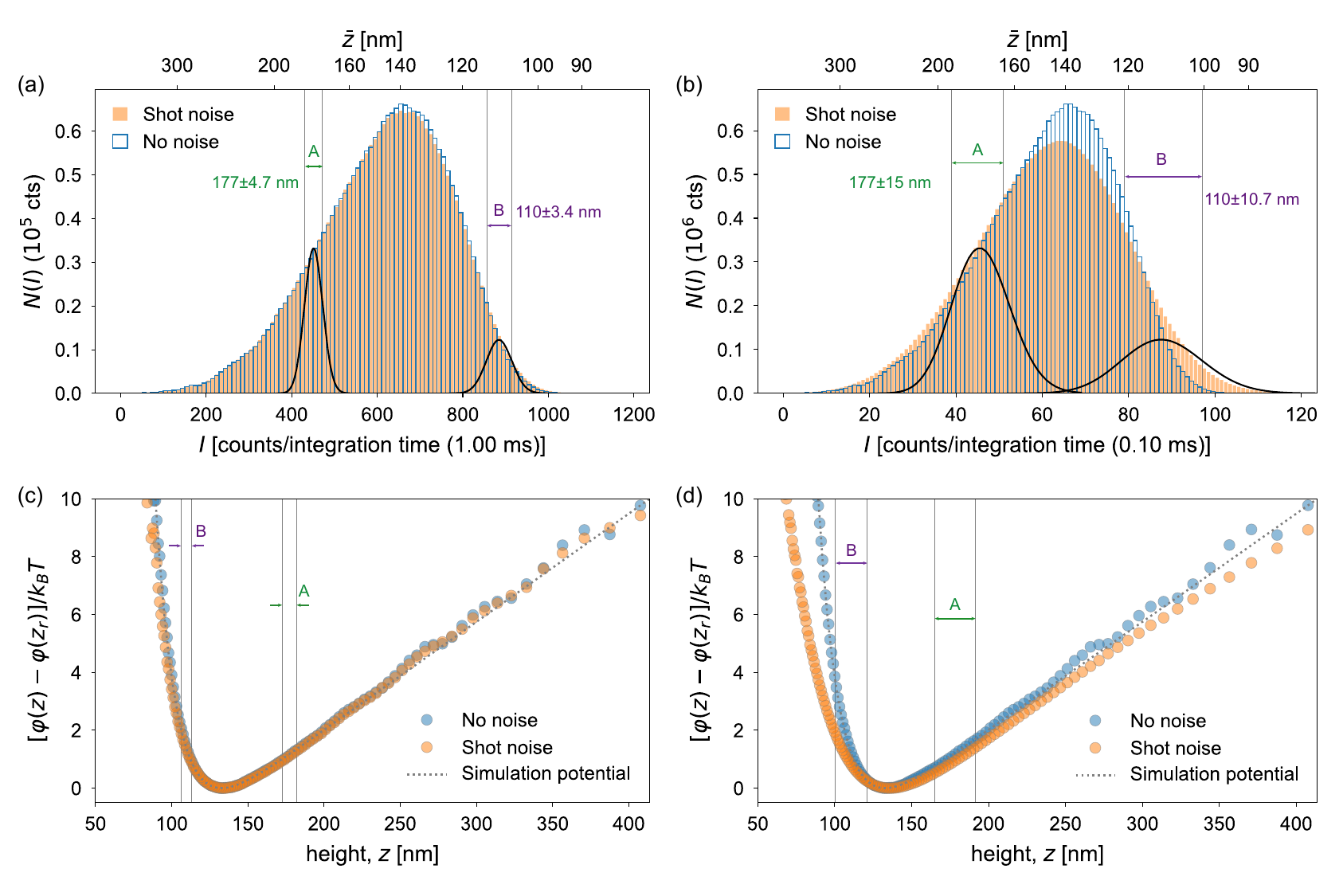}
        \caption{
            \label{fig:0.5mMYukawa}
            Scattering intensity histograms from a simulation of an 8.0-$\mu$m-diameter PS sphere in 0.5 mM monovalent saline solution with a photon counting time interval $\tau$ of
            (a) 1 ms and
            (b) 0.1 ms.
            The solid orange and empty blue bars show histograms $P(n)$ and $P(n_c)$ obtained with and without shot noise included, respectively.
            The solid black curves show the Poisson distributions when the means in panel (a) are $\bar{n} = $450 and 880 counts and in panel (b) are $\bar{n} = $45 and 88 counts.
            The widths of the Poisson distributions are indicated by vertical gray lines drawn at $n=\bar{n} \pm \sqrt{\bar{n}}$ in panels (a) and (b).
            The potential curves are calculated with $\tau$ equal to
            (c) 1 ms and (d) 0.1 ms.
            The solid orange and empty blue circles show potential curves obtained with and without shot noise included, respectively.
            The dotted lines show the potential curve used in the simulation.
        }
    \end{figure*}

    \subsection{Effect of shot noise on double-layer repulsion and gravity potentials}

    To evaluate the effect of shot noise on the potential profile obtained from TIRM measurements, we simulate the scattering intensities using integration times $\tau$ of 1 and 0.1 ms.
    For each value of $\tau$, we construct histograms of the simulated scattering intensities in two ways, one with shot noise, $P(n)$, and the other without, $P(n_c)$.
    The results are shown in Fig.\ \ref{fig:0.5mMYukawa}(a) and (b) for $\tau$ of 1 and 0.1 ms, respectively.

    We then calculate the interaction potentials for the two cases, and compare them with analytical predictions from eqn \eqref{eq:G-DLpotential}, as shown in Fig.\ \ref{fig:0.5mMYukawa}(c) and (d).
    When $\tau = 1$~ms, the simulated potential shown in Fig.\ \ref{fig:0.5mMYukawa}(c) with and without shot noise are both in good agreement with the ideal analytical profile. The fitted Debye length from simulations with and without shot noise are 15.5 nm and 13.9 nm; while the fitted buoyant weights $G$ are 0.156 pN and 0.157, respectively.
    Both are reasonably close to the true values of Debye length (13.7 nm) and buoyant weight (0.152 pN) we input for the simulation.
    Indeed, this conclusion could be anticipated from Fig.\ \ref{fig:0.5mMYukawa}(a), where the two histograms with and without shot noise, $P(n)$ and $P(n_c)$, are nearly indistinguishable.

    On the other hand, the effect of shot noise becomes significant when a smaller value of $\tau=0.1$~ms is used.
    As shown in Fig.\ \ref{fig:0.5mMYukawa}d, the potential curve with shot noise is visibly broadened.
    Notably, the distortion of the potential profile in Fig.\ \ref{fig:0.5mMYukawa}(d) is much more pronounced in the short-range electrostatic repulsion region than in the gravity-dominated region.
    The broadened shape yields a Debye length of 23.6~nm, significantly longer than the true Debye length of 13.7~nm.
    The fitted buoyant weight, $G = 0.157$~pN, is however close to the true value of 0.152~pN.

    \begin{figure*}[h!]
        \centering
        \includegraphics{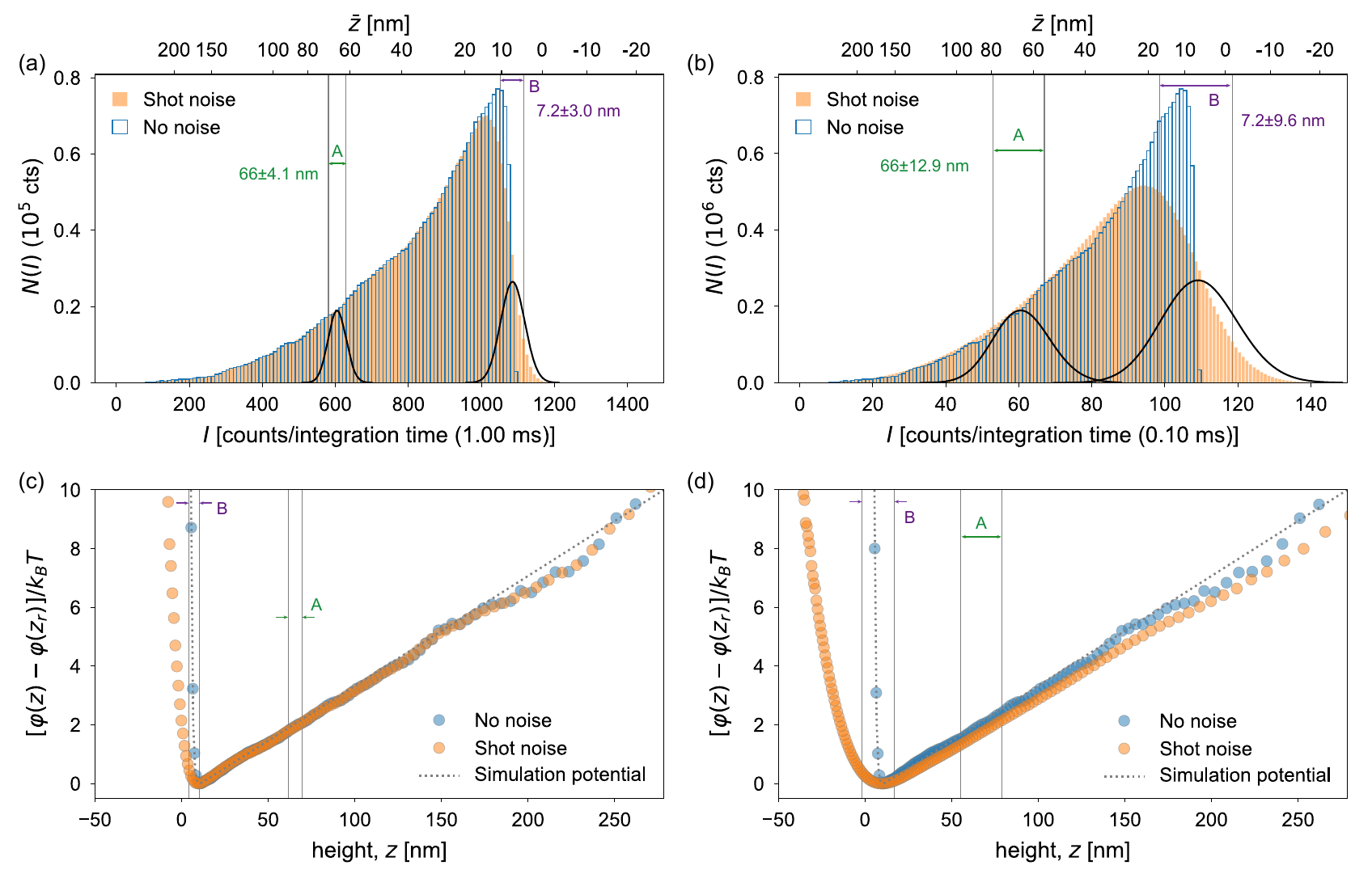}
        \caption{
            \label{fig:140mMYukawa}
            Scattering intensity histograms from a simulation of an 8.0-$\mu$m-diameter PS sphere in 140 mM monovalent saline solution with a photon counting time interval $\tau$ of
            (a) 1 ms and
            (b) 0.1 ms.
            The solid orange and empty blue bars show histograms $P(n)$ and $P(n_c)$ obtained with and without shot noise included, respectively.
            The solid black curves show the Poisson distributions when the means in panel (a) are $\bar{n} =$ 600 and 1080 counts and in panel (b) are $\bar{n} =$ 60 and 108 counts.
            The widths of the Poisson distributions are indicated by vertical gray lines drawn at $n=\bar{n} \pm \sqrt{\bar{n}}$ in panels (a) and (b).
            The potential curves are calculated with $\tau$ equal to
            (c) 1 ms and (d) 0.1 ms.
            The solid orange and empty blue circles show potential curves obtained with and without shot noise included, respectively.
            The dotted lines show the potential curve used in the simulation.
        }
    \end{figure*}

    Two factors play important roles in determining the degree of distortion of the measured intensity histogram $P(n)$ and the potential $\varphi(z)$ derived from $P(n)$: (1) the number of counts $n$ per integration time $\tau$ and (2) the steepness (spatial gradient $\partial\varphi/\partial z$) of the potential.

    In Fig.\ \ref{fig:0.5mMYukawa}a and b, we plot the intensity axis as the number of counts $n$ per integration time $\tau$ in order to highlight the role of $n$ in determining the shot noise.
    Because the maximum count rate (intensity) is the same in both cases, approximately $10^6$~cts/s, the number of counts per integration time is a factor of 10 lower for the case where $\tau=0.10$~ms compared to the case where $\tau=1.00$~ms.
    If there were no shot noise associated with photon counting, both data sets would result in statistically the same histogram of intensities, because in this hypothetical case there would be a one-to-one correspondence between the number of counts per integration time and particle height $z$, as given by eqn \eqref{eq:n_c}.
    Changing the number of counts per integration time simply changes the horizontal intensity scale of the histogram but nothing else.
    This is what is shown by the histograms of open blue bars in in Fig.\ \ref{fig:0.5mMYukawa}a and b, which correspond to the probability distribution $P(n_c)$ introduced in eqn \eqref{eq:Pcon}.
    Indeed, the potentials $\varphi(z)$ determined from these two histograms both track the expected result, as shown by the blue data points in Fig.\ \ref{fig:0.5mMYukawa}c and d, which follow the dotted lines indicating the theoretically-defined potential that was used in the Brownian dynamics simulation.

    \begin{figure*}[h!]
        \includegraphics{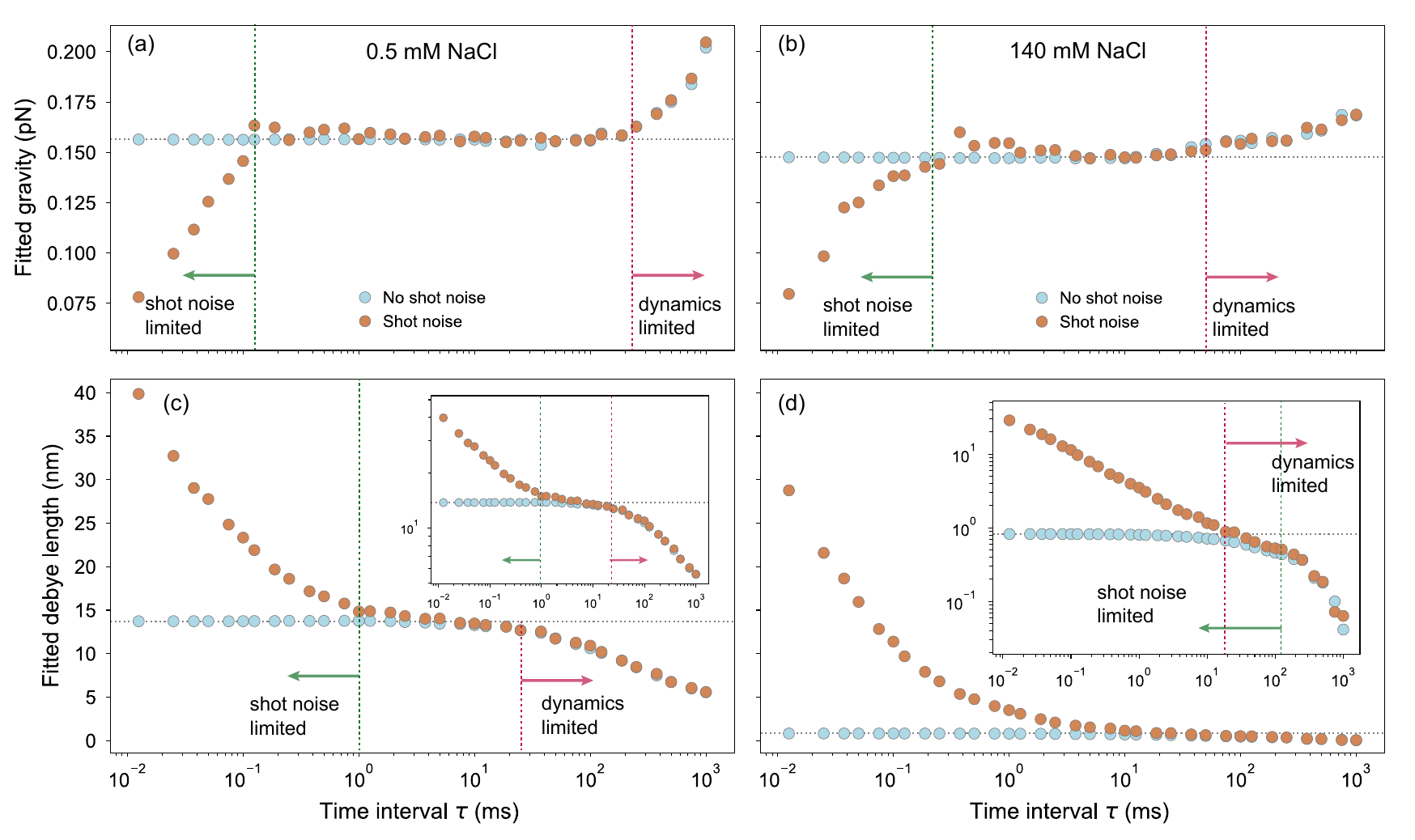}
        \caption{
            \label{fig:Yukawa_parameter}
            Simulated interaction parameters of an 8-$\mu$m-diameter PS sphere in (a), (c) 0.5 mM and (b), (d) 140 mM monovalent salt.
            (a), (b) Fitted buoyant weight $G$ and
            (c), (d) Debye length $\kappa^{-1}$ from potential profiles simulated with different $\tau$.
            Blue circle are fitted parameters without considering of photon counting uncertainties.
            Dotted lines are the theoretical values of the expected parameter.
            (d) inset is the same plot in (d) shown in y-log scale.
        }
    \end{figure*}

    To emphasize the one-to-one correspondence between the average counts $n_c$ and $z$, which are related by eqn \eqref{eq:n_c}, we include a second horizontal axis in Fig.\ \ref{fig:0.5mMYukawa}a and b that gives the values of $z$ associated with $n_c$ axis.

    Next we consider the actual case where there is shot noise from photon counting.
    The histograms obtained in this case are shown as solid orange bars in Fig.\ \ref{fig:0.5mMYukawa}a and b and correspond the $P(n)$, which is what is measured experimentally.
    In this case the histograms for the two integration times of 1.0~ms and 0.1~ms differ dramatically, particularly for the higher intensities in each plot, which correspond to the smaller values of $z$ where the potential $\varphi(z)$ is steepest.
    The differences in the histograms obtained with and without photon counting shot noise can be understood by recalling eqn \eqref{eq:Pcon}:
    \begin{align*}
         P(n) = \sum_{n_c} P(n_c)\, P_{\tau}(n; n_c) \;,
    \end{align*}
    where $P_{\tau}(n; n_c)$ is the Poisson distribution given by eqn \eqref{eq:photondist}.
    In Fig.\ \ref{fig:0.5mMYukawa}a and b, we plot $P_{\tau}(n; n_c)$ on top of the histograms for the values of $n_c$ corresponding to two values of $z$: 177~nm and 110~nm.
    These two values correspond to two different sets of values for $n$: 450 and 880 for Fig.\ \ref{fig:0.5mMYukawa}a, where $\tau=1.0$~ms, and 45 and 88 for Fig.\ \ref{fig:0.5mMYukawa}b, where $\tau=0.1$~ms.

    From Fig.\ \ref{fig:0.5mMYukawa}a and b, we see that for a given value of $n$, $P(n)$ is smeared out over the range spanned by $P_{\tau}(n;n_c)$, or about $\pm \sqrt{n_c}$.
    Note, however, that what matters when comparing the histograms in Fig.\ \ref{fig:0.5mMYukawa}a and b is not the width of  $P_{\tau}(n;n_c)$ but the \textit{relative} width, which is $\pm \sqrt{n_c}/n_c = \pm n_c^{-1/2}$.
    Thus, as illustrated by Fig.\ \ref{fig:0.5mMYukawa}a and b, the rounding effect of shot noise on $P(n)$ is greater when the counts per integration time is smaller.

    The other feature that plays an important role in the distortion of $P(n)$ by shot noise is the gradient of the potential.
    A steep potential leads to a steep $P(n_c)$, which in turn is more readily rounded by the shot noise distribution $P_{\tau}(n;n_c)$.
    This is evident in Fig.\ \ref{fig:0.5mMYukawa}b where the width of the noise distribution $P_{\tau}(n;n_c)$ is comparable to the width of the change in $P(n_c)$ for $n \sim 88$.

    For each plot, we pick out two values of particle height $\bar{z}$, 110~nm and 177~nm, and show using solid black lines the Poisson distribution for the corresponding value of $n_c$ from eqn \eqref{eq:photondist}.
    These curves give the distribution of values of $n$ that are measured in a TIRM experiment for a particle at a single height $z$.
    For example, in Fig.\ \ref{fig:0.5mMYukawa}a, for $\bar{z}=177$~nm (label A), $n_c=450$~cts per integration time, for which the Poisson distribution has a width of $\sqrt{n_c} = \sqrt{450} \simeq 21$.
    According to eqn \eqref{eq:error}, this leads to an uncertainty in $z$ of $\pm 4.7$~nm, where $\beta^{-1}=100$~nm.
    Similarly, in Fig.\ \ref{fig:0.5mMYukawa}b, for $\bar{z}=177$~nm (label A), $n_c=45$~cts per integration time, for which the Poisson distribution has a width of $\sqrt{n_c} = \sqrt{45} \simeq 6.7$.
    According to eqn \eqref{eq:error}, this leads to an uncertainty in $z$ of $\pm 15$~nm, where $\beta^{-1}=100$~nm.
    The relative uncertainty $\Delta n_c/n_c = n_c^{-1/2}$ is much larger when $n_c$ is small, which is also reflected in the relative widths of the Poisson distributions in Fig.\ \ref{fig:0.5mMYukawa}a and b.

    Performing the same analysis for $\bar{z}=110$~nm (label B), where $n_c = 880$ and 88~cts per integration time, respectively, in Fig.\ \ref{fig:0.5mMYukawa}a and b, we obtain uncertainties in $z$ of 3.4~nm and 10.7~nm for the two cases.

    The uncertainty in position leads to an uncertainty in the potential, which can be roughly estimated by
    \begin{align}
        \Delta\varphi(n)
        = \left(\frac{\partial \varphi }{\partial z}\right) \left(\frac{\partial z }{\partial n}\right) \Delta n
        %= -\left(\frac{\partial \varphi }{\partial z}\right) \frac{\beta^{-1}}{ n} \Delta n
        = -\left(\frac{\partial \varphi }{\partial z}\right) \frac{\beta^{-1}}{\sqrt{n}} \;.
        \label{eq:phi_error}
    \end{align}
    where $n$ is the number of scattering photons counted by the detector within the integration time $\tau$ (i.e., $n$ = $I\tau$), and thus the photon shot noise is $\sqrt{n}$.
    One can see that the potential gradient $\partial \varphi/ \partial z$, or the force, plays a critical role in determining how strongly photon shot noise can affect the derived potential.
    For example, in the gravity-dominated region (location B) with $\tau$ = 0.1~ms where the potential changes very slowly, the height uncertainty is as large as 26.8 nm, but the noise-corrupted potential curve still shows good agreement with the theoretical prediction.
    By contrast, in the sharp electrostatic repulsion region (location B), the noise-corrupted intensities do not provide a faithful representation of the potential shape, leading to the broadening of the resultant potential profile.

    To further demonstrate the effect of potential sharpness on measurement tolerance against photon shot noise, we increase the salt concentration to 140 mM in the simulation, creating a much sharper potential in the double-layer repulsion region with Debye length of 0.82 nm.
    Fig.\ \ref{fig:140mMYukawa} shows the effect of shot noise on intensity histograms when $\tau$ is 1 ms and 0.1 ms in panels (a) and (b), respectively.
    In this case, the potential is so sharp that neither choice of integration times provides a completely faithful measurement of the interaction potential.
    Nevertheless, using the shorter integration time of $\tau = 0.1$ ms yields significantly poorer results than using $\tau = 1$ ms, as shown in Fig.\ \ref{fig:140mMYukawa}c and d.

    In Fig.\ \ref{fig:Yukawa_parameter}, we summarize the fitting parameters extracted from our simulated TIRM data for integration time intervals spanning the range from $\tau=0.025$~ms to 1000~ms for the two monovalent salt concentrations considered above: 0.5~mM, which gives a soft repulsive potential with  $\kappa^{-1}=13.7$~nm, and 140~mM, which gives a hard repulsive potential with $\kappa^{-1}=0.82$~nm.

    Fig.\ \ref{fig:Yukawa_parameter}a and b show that the buoyant weight $G$ is well fit over a broad range of $\tau$, which simply reflects the fact that the potential does not vary sharply in the large-distance ($z \gtrsim 30$~nm) gravity-dominated part of the potential from which the fitted value of $G$ is extracted.

    By contrast, Fig.\ \ref{fig:Yukawa_parameter}c shows that the Debye length $\kappa^{-1}$ is well fit only over a relatively narrow range, $1~\mathrm{ms} < \tau < 25~\mathrm{ms}$ for a soft repulsive potential with $\kappa^{-1} = 13.7$~nm while Fig.\ \ref{fig:Yukawa_parameter}c shows that $\kappa^{-1}$ is not well fit at all for a hard repulsive potential with $\kappa^{-1} = 0.82$~nm, except near $\tau = 30$~ms where different offsetting errors, which we discuss next, accidentally cancel.

    The errors in the fitted values of $G$ and $\kappa^{-1}$ for the smaller values of $\tau$ arise from the photon counting shot noise which broadens the intensity distribution $N(I)$ and thus broadens the potential.
    This decreases the fitted values of $G$ and increases the fitted values of $\kappa^{-1}$.
    The errors in the fitted values of $G$ and $\kappa^{-1}$ for the larger values of $\tau$ arise from particles diffusing too far---particle dynamics---which suppresses the wings of the intensity distribution $N(I)$ making the potential that is inferred from it sharper, which increases the fitted value of $G$ and decreases the fitted value of $\kappa^{-1}$.

    As a check on our simulations, potentials extracted from both simulated and real experimental data are compared for the case $\kappa^{-1}=13.7$~nm in Fig.\ S5 in the Supplemental Information for a wide range of integration times $\tau$.
    We find that the simulated data sets agree very well with real experimental data and are fully consistent with the fitting parameters shown in Fig.\ \ref{fig:Yukawa_parameter}a and c.

    The data sets analyzed in Fig.\ \ref{fig:Yukawa_parameter} serve as a cautionary note for interpreting the potentials obtained in TIRM experiments.
    Thus, one may ask how to determine if the potentials obtained using TIRM are artificially broadened or narrowed.
    A simple way to check if the photon shot noise is broadening the potentials is to plot the Poisson distributions on the intensity histograms and compare their width to the slope of $N(I)$ as we have done in Figs.\ \ref{fig:0.5mMYukawa}a and c and \ref{fig:140mMYukawa}a and c.
    While this is helpful in determining whether $\tau$ is too small such that shot noise is broadening $N(I)$, it does not aid in determining if $\tau$ is too long so that particle dynamics are narrowing $N(I)$.
    The best results are obtained when these two regimes of shot noise and dynamics-limited data are well separated, as they are in Fig.\ \ref{fig:Yukawa_parameter}a and c when the potential is relatively soft.
    In this case, there is a range of values of $\tau$ that give nearly identical results; here that occurs for $1~\mathrm{ms} < \tau < 25~\mathrm{ms}$.

    \begin{figure}[h!]
        \includegraphics{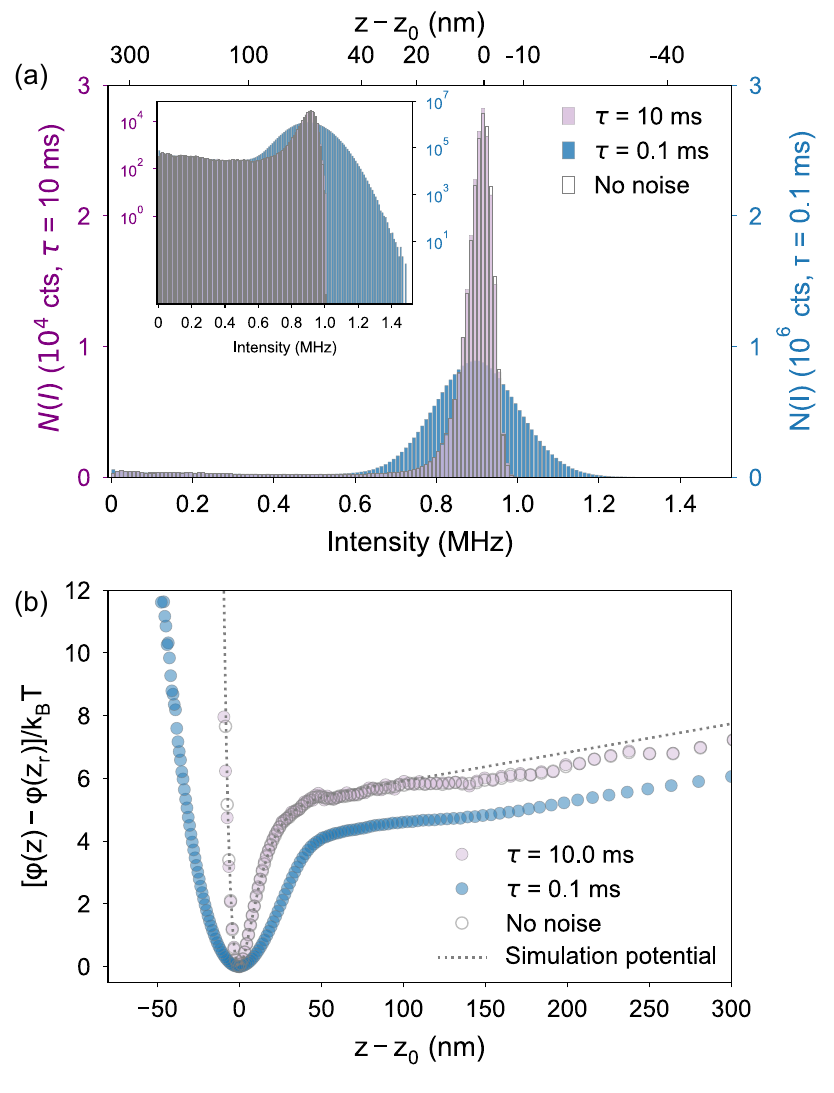}
        \caption{
            \label{fig:Morsepotential}
            (a) Histogram of simulated scattered intensities of a 5-$\mu$m-diameter PS particle with Morse potential profile($\epsilon = 5 k_BT$, $a$ = 10 nm, $z_0$ = $a$.).
            Step size is 5 $\mu$s with $5\!\times\!10^8$ steps.
            $I_{max}$ = 1 MHz.
            Grey bars are intensities from original simulated trajectories without account for shot noise or photon counting interval (effectively $\tau$ = $\Delta$t).
            Blue, purple bars correspond to intensities with $\tau$ being 0.1 ms and 10 ms respectively. Inset shows the same plot in y-log scale.
            (b) Potential curves corresponding to the conditions in (a) when $\tau$ = 0.1 ms (blue), 10 ms (purple) and 5 $\mu$s (grey, no shot noise).
            Potential minimum is placed at $0~k_BT$ and height at 0 nm.
        }
    \end{figure}

    \subsection{Effect of shot noise on measurement of potential wells: Morse potential}

    In this section, we investigate the effect of shot noise on the measurement of a short-range attractive interaction.
    Such potentials are common in interacting particle systems and include the depletion interaction\cite{rudhardt1998direct}, attractive electrostatic\cite{leunissen2005vanBlaadIonic,sacanna2020ionicTed}, Casimir\cite{bechinger2008casimirTIRM}, and the interaction between DNA-coated colloids \cite{RogersPNAS,rogers2021dnaccMembrane}.
    The range of these attractions goes from about a nanometer to hundreds of nanometers.

    To model a generic attractive interaction, we choose a Morse potential as our model potential, as it is frequently used to describe attractive interactions between colloidal particles \cite{doye1995effect,bou2020sticky}:
    \begin{align}
        \varphi(z) = \epsilon [e^{-2(z-z_0)/a} - 2e^{-(z-z_0)/a} ] \;.
        \label{eq:Morse_potential}
    \end{align}
    It has a depth of $\epsilon$ that occurs at a height $z_0$; the width and stiffness of the potential well are conveniently set by $a$.

    Consider a colloidal sphere that has close-range attractive interactions with the glass surface.
    Its potential energy $\varphi(z)$ can be written a combination of the Morse potential and gravitation:
    \begin{align}
        \varphi(z) = \epsilon [e^{-2(z-z_0)/a} - 2e^{-(z-z_0)/a} ] + Gz \;.
        \label{eq:Morse_potential_G}
    \end{align}

    For TIRM, we need to choose a reference height $z_r$, which we take to be $z_0$.
    With this choice,
    \begin{align}
        \varphi(z) - \varphi(z_0) = \epsilon [e^{-2(z-z_0)/a} - 2e^{-(z-z_0)/a} + 1] + G(z-z_0),
    \end{align}

    \begin{figure}[t!]
        \includegraphics{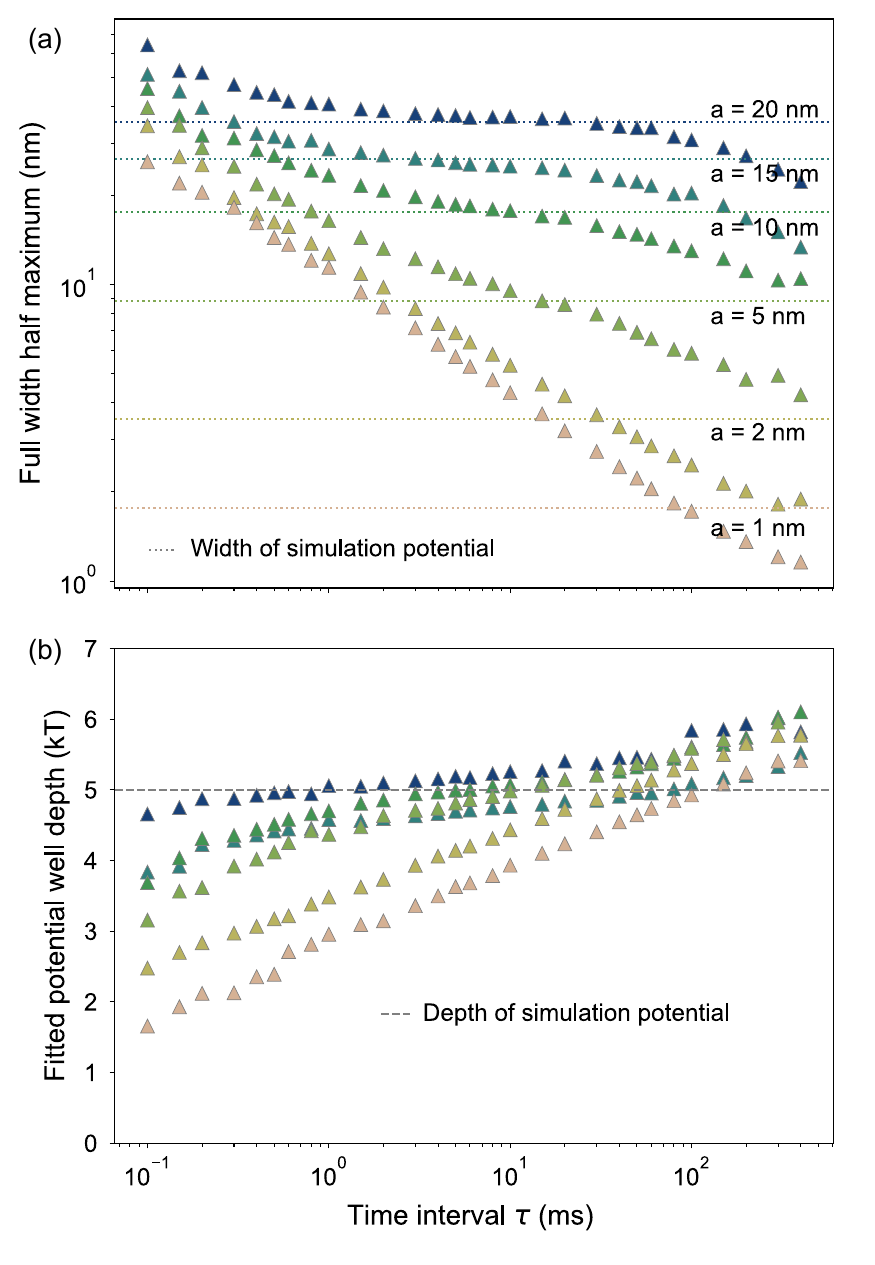}
        \caption{
            \label{fig:MorseParameter}  Simulated $5~k_BT$ deep Morse potentials with different $a$ and the corresponding apparent potential width (FWHM)
            (a) and depth
            (b) when using different $\tau$.
            Horizontal colored dotted lines in (a) and gray dashed line in (b) indicate the theoretical width or depth for the Morse potentials.
        }
    \end{figure}

    Using the Brownian dynamics simulation method introduced in \S\ref{sec:sim}, we simulate the height trajectories of an 5-$\mu$m PS particle with a Morse potential energy profile.
    Fig.\ \ref{fig:Morsepotential} shows the Morse potential simulation results with $a$ = 10 nm and $\epsilon = 5 k_BT$.
    Equation \eqref{eq:Morse_potential_G} is plotted as a gray dotted line in Fig.\ \ref{fig:Morsepotential}.

    We carry out simulations using different photon-counting integration time intervals $\tau$.
    Fig.\ \ref{fig:Morsepotential} shows results for two values of $\tau$: 0.1 ms and 10 ms.

    When $\tau = 10$~ms, the light scattering intensity distribution $N(I)$ is essentially indistinguishable from what one would obtain in the absence of photon counting shot noise, as can be seen in Fig.\ \ref{fig:Morsepotential}a.
    In this case, the potential obtained from the usual TIRM analysis corresponds closely to the true potential, as can be seen in Fig.\ \ref{fig:Morsepotential}b.

    When $\tau$ is reduced to 0.1~ms, however, the intensity distribution $N(I)$ is blurred considerably, particularly when the particle is in the vicinity of the potential minimum, as shown in Fig.\ \ref{fig:Morsepotential}a.
    As a consequence, the potential obtained from the TIRM analysis is broadened considerably.
    The depth of the potential is also reduced.

    To better quantify the effect of $\tau$ on potential well measurements, we simulate Morse potentials with varying width, taking $a$ as 1, 2, 5, 10, 15, and 20 nm.
    For each value of $a$, we use different values $\tau$ to simulate different TIRM data sets from which we extract a measured potential.
    We characterize the potential well inferred from a TIRM measurement by two parameters: the well depth and width.
    To characterize the well depth and width, we first subtract the contribution from gravity, $G(z-z_0)$, from the potential (see Fig. S6 in SI).
    We define the characteristic width as the full width at half maximum (FWHM).

    Fig.\ \ref{fig:MorseParameter}a shows the potential width (FWHM) measured from the simulated TIRM data as a function of $\tau$ for different values of $a$.
    %The gray horizontal dotted lines indicate the simulated $\mathrm{FWHM}$ for each value of the width parameter $a$ without the influence of the integration time and shot noise.
    The gray horizontal dotted lines indicate the true $\mathrm{FWHM} = 1.763a$ for each value of the width parameter $a$.
    For each of the wider Morse potentials with $a = 20$, 15, and 10 nm, there is a clear plateau over a range of integration times $\tau$ near the expected width.
    One can discern evidence of an incipient plateau for $a=5$~nm but all evidence of a plateau vanishes for the smaller values of $a$.
    Nevertheless, even below $a=5$~nm, the measured width is different for different values of $a$ for a given choice of $\tau$, indicating that even in this limit TIRM can be expected to follow \textit{changes} in the width of the potential.

    Fig.\ \ref{fig:MorseParameter}b shows the simulated potential depth obtained using different values of $\tau$.
    Here we see that the correct potential depth is obtained only for a somewhat narrower range of $\tau$ centered around a few milliseconds, where there is a mild plateau, again for wells wider than about 5 nm.
    Thus, correct measurements of the well depth and width can be obtained, but only for a somewhat narrow range of $\tau$.
    The proper range of $\tau$ can be identified by the existence of a plateau.

    \section{Conclusions}
    In this paper, we establish the critical role of shot noise and intensity integration time $\tau$ in TIRM measurements of colloidal interaction potentials. Sharp potential profiles, e.g. steric repulsion \cite{prlang2007steric}, depletion \cite{bevan2012tirmdepletion}, strong double-layer repulsion\cite{noisevdwnayeri2013} and close-range surface bindings\cite{protein--carbohydrateBevan2013,rogers2021dnaccMembrane} are particular prone to the corruption of shot noise. Shot noise should be taken into consideration when interpreting such measured potential profiles. While shot noise cannot be removed from deconvolution, the expected signal-to-noise ratio can be experimentally controlled by selecting the photon counting time interval.

    The choice of photon counting time interval is a trade-off between minimizing shot noise and preserve the temporal resolution.
    For smaller $\tau$,  the temporal resolution of the simulated measurement is sufficient to resolve the particle movement.
    However, small $\tau$ means large shot noise and tends to blur the sharp features of a potential curve.
    For large $\tau$, shot noise is reduced due to the increased number of photons counted.
    However, the measured scattering intensity is averaged over large particle displacements, which will distort the potential profiles for excessively large $\tau$.
    If spatial gradients $\partial \varphi/\partial z$ in the potential are not too large, there can be a regime of intermediate values of $\tau$ for which the intensity distribution $N(I)$ is accurately measured without significant distortion.
    In this case, the potential $\varphi(z)$ can be faithfully measured by TIRM.
    Even in cases where the photon counting shot noise cannot be reduced to the point that it does not broaden sharp features in $\varphi(z)$, useful information can be extracted about the potential and, in particular, changes in $\varphi(z)$ due to changing sample conditions (temperature, salt concentration, \textit{etc.}) can be readily discerned.

    \section*{Conflicts of interest}
    There are no conflicts to declare.

    \section*{Acknowledgments}
    This research was primarily supported by Department of Energy (DOE) DE-SC0007991 for the initiation and design of the experiments.
    Additional funding was supplied by the US Army Research Office under award number W911NF-17-1-0328.

    %%%END OF MAIN TEXT%%%

    %The \balance command can be used to balance the columns on the final page if desired. It should be placed anywhere within the first column of the last page.

    \balance

    %If notes are included in your references you can change the title from 'References' to 'Notes and references' using the following command:
    %\renewcommand\refname{Notes and references}

    %%%REFERENCES%%%
    \bibliography{tirmshot} %You need to replace "rsc" on this line with the name of your .bib file

\providecommand*{\mcitethebibliography}{\thebibliography}
\csname @ifundefined\endcsname{endmcitethebibliography}
{\let\endmcitethebibliography\endthebibliography}{}
\begin{mcitethebibliography}{35}
\providecommand*{\natexlab}[1]{#1}
\providecommand*{\mciteSetBstSublistMode}[1]{}
\providecommand*{\mciteSetBstMaxWidthForm}[2]{}
\providecommand*{\mciteBstWouldAddEndPuncttrue}
  {\def\EndOfBibitem{\unskip.}}
\providecommand*{\mciteBstWouldAddEndPunctfalse}
  {\let\EndOfBibitem\relax}
\providecommand*{\mciteSetBstMidEndSepPunct}[3]{}
\providecommand*{\mciteSetBstSublistLabelBeginEnd}[3]{}
\providecommand*{\EndOfBibitem}{}
\mciteSetBstSublistMode{f}
\mciteSetBstMaxWidthForm{subitem}
{(\emph{\alph{mcitesubitemcount}})}
\mciteSetBstSublistLabelBeginEnd{\mcitemaxwidthsubitemform\space}
{\relax}{\relax}

\bibitem[Prieve(1999)]{prieve1999tirm}
D.~C. Prieve, \emph{Adv. Colloid Interface Sci.}, 1999, \textbf{82},
  93--125\relax
\mciteBstWouldAddEndPuncttrue
\mciteSetBstMidEndSepPunct{\mcitedefaultmidpunct}
{\mcitedefaultendpunct}{\mcitedefaultseppunct}\relax
\EndOfBibitem
\bibitem[Prieve and Frej(1990)]{prieve1990tirm}
D.~C. Prieve and N.~A. Frej, \emph{Langmuir}, 1990, \textbf{6}, 396--403\relax
\mciteBstWouldAddEndPuncttrue
\mciteSetBstMidEndSepPunct{\mcitedefaultmidpunct}
{\mcitedefaultendpunct}{\mcitedefaultseppunct}\relax
\EndOfBibitem
\bibitem[Bike(2000)]{bike2000tirm}
S.~G. Bike, \emph{Curr. Opin. Colloid Interface Sci.}, 2000, \textbf{5},
  144--150\relax
\mciteBstWouldAddEndPuncttrue
\mciteSetBstMidEndSepPunct{\mcitedefaultmidpunct}
{\mcitedefaultendpunct}{\mcitedefaultseppunct}\relax
\EndOfBibitem
\bibitem[Bevan and Prieve(2000)]{prieve2000tirmbevan}
M.~A. Bevan and D.~C. Prieve, \emph{Langmuir}, 2000, \textbf{16},
  9274--9281\relax
\mciteBstWouldAddEndPuncttrue
\mciteSetBstMidEndSepPunct{\mcitedefaultmidpunct}
{\mcitedefaultendpunct}{\mcitedefaultseppunct}\relax
\EndOfBibitem
\bibitem[Gong \emph{et~al.}(2014)Gong, Wang, and Ngai]{ngai2014tirm}
X.~Gong, Z.~Wang and T.~Ngai, \emph{Chem. Commun.}, 2014, \textbf{50},
  6556--6570\relax
\mciteBstWouldAddEndPuncttrue
\mciteSetBstMidEndSepPunct{\mcitedefaultmidpunct}
{\mcitedefaultendpunct}{\mcitedefaultseppunct}\relax
\EndOfBibitem
\bibitem[Kleshchanok and Lang(2007)]{prlang2007steric}
D.~Kleshchanok and P.~R. Lang, \emph{Langmuir}, 2007, \textbf{23},
  4332--4339\relax
\mciteBstWouldAddEndPuncttrue
\mciteSetBstMidEndSepPunct{\mcitedefaultmidpunct}
{\mcitedefaultendpunct}{\mcitedefaultseppunct}\relax
\EndOfBibitem
\bibitem[Edwards and Bevan(2012)]{bevan2012tirmdepletion}
T.~D. Edwards and M.~A. Bevan, \emph{Langmuir}, 2012, \textbf{28},
  13816--13823\relax
\mciteBstWouldAddEndPuncttrue
\mciteSetBstMidEndSepPunct{\mcitedefaultmidpunct}
{\mcitedefaultendpunct}{\mcitedefaultseppunct}\relax
\EndOfBibitem
\bibitem[July \emph{et~al.}(2012)July, Kleshchanok, and Lang]{depletionprlang}
C.~July, D.~Kleshchanok and P.~Lang, \emph{Eur. Phys. J. E}, 2012, \textbf{35},
  1--8\relax
\mciteBstWouldAddEndPuncttrue
\mciteSetBstMidEndSepPunct{\mcitedefaultmidpunct}
{\mcitedefaultendpunct}{\mcitedefaultseppunct}\relax
\EndOfBibitem
\bibitem[Hertlein \emph{et~al.}(2008)Hertlein, Helden, Gambassi, Dietrich, and
  Bechinger]{bechinger2008tirmcasimir}
C.~Hertlein, L.~Helden, A.~Gambassi, S.~Dietrich and C.~Bechinger,
  \emph{Nature}, 2008, \textbf{451}, 172--175\relax
\mciteBstWouldAddEndPuncttrue
\mciteSetBstMidEndSepPunct{\mcitedefaultmidpunct}
{\mcitedefaultendpunct}{\mcitedefaultseppunct}\relax
\EndOfBibitem
\bibitem[Merminod \emph{et~al.}(2021)Merminod, Edison, Fang, Hagan, and
  Rogers]{rogers2021dnaccMembrane}
S.~Merminod, J.~R. Edison, H.~Fang, M.~F. Hagan and W.~B. Rogers,
  \emph{Nanoscale}, 2021, \textbf{13}, 12602--12612\relax
\mciteBstWouldAddEndPuncttrue
\mciteSetBstMidEndSepPunct{\mcitedefaultmidpunct}
{\mcitedefaultendpunct}{\mcitedefaultseppunct}\relax
\EndOfBibitem
\bibitem[Clapp \emph{et~al.}(1999)Clapp, Ruta, and Dickinson]{clapp1999noise}
A.~R. Clapp, A.~G. Ruta and R.~B. Dickinson, \emph{Rev. Sci. Inst.}, 1999,
  \textbf{70}, 2627--2636\relax
\mciteBstWouldAddEndPuncttrue
\mciteSetBstMidEndSepPunct{\mcitedefaultmidpunct}
{\mcitedefaultendpunct}{\mcitedefaultseppunct}\relax
\EndOfBibitem
\bibitem[Clapp and Dickinson(2001)]{clapp2001noise}
A.~R. Clapp and R.~B. Dickinson, \emph{Langmuir}, 2001, \textbf{17},
  2182--2191\relax
\mciteBstWouldAddEndPuncttrue
\mciteSetBstMidEndSepPunct{\mcitedefaultmidpunct}
{\mcitedefaultendpunct}{\mcitedefaultseppunct}\relax
\EndOfBibitem
\bibitem[Helseth and Fischer(2004)]{helseth2004fundamental}
L.~Helseth and T.~M. Fischer, \emph{J. Colloid Interface Sci.}, 2004,
  \textbf{275}, 322--327\relax
\mciteBstWouldAddEndPuncttrue
\mciteSetBstMidEndSepPunct{\mcitedefaultmidpunct}
{\mcitedefaultendpunct}{\mcitedefaultseppunct}\relax
\EndOfBibitem
\bibitem[Frej and Prieve(1993)]{frej1993hindered}
N.~A. Frej and D.~C. Prieve, \emph{J. Chem. Phys.}, 1993, \textbf{98},
  7552--7564\relax
\mciteBstWouldAddEndPuncttrue
\mciteSetBstMidEndSepPunct{\mcitedefaultmidpunct}
{\mcitedefaultendpunct}{\mcitedefaultseppunct}\relax
\EndOfBibitem
\bibitem[Oetama and Walz(2005)]{waltzmentionshot2005new}
R.~J. Oetama and J.~Y. Walz, \emph{J. Colloid Interface Sci.}, 2005,
  \textbf{284}, 323--331\relax
\mciteBstWouldAddEndPuncttrue
\mciteSetBstMidEndSepPunct{\mcitedefaultmidpunct}
{\mcitedefaultendpunct}{\mcitedefaultseppunct}\relax
\EndOfBibitem
\bibitem[Chew \emph{et~al.}(1979)Chew, Wang, and Kerker]{kerker79evanescent}
H.~Chew, D.-S. Wang and M.~Kerker, \emph{Appl. Opt.}, 1979, \textbf{18},
  2679--2679\relax
\mciteBstWouldAddEndPuncttrue
\mciteSetBstMidEndSepPunct{\mcitedefaultmidpunct}
{\mcitedefaultendpunct}{\mcitedefaultseppunct}\relax
\EndOfBibitem
\bibitem[Helden \emph{et~al.}(2006)Helden, Eremina, Riefler, Hertlein,
  Bechinger, Eremin, and Wriedt]{helden2006single}
L.~Helden, E.~Eremina, N.~Riefler, C.~Hertlein, C.~Bechinger, Y.~Eremin and
  T.~Wriedt, \emph{Appl. Opt.}, 2006, \textbf{45}, 7299--7308\relax
\mciteBstWouldAddEndPuncttrue
\mciteSetBstMidEndSepPunct{\mcitedefaultmidpunct}
{\mcitedefaultendpunct}{\mcitedefaultseppunct}\relax
\EndOfBibitem
\bibitem[Odiachi and Prieve(2004)]{prieve_remove_noise}
P.~C. Odiachi and D.~C. Prieve, \emph{J. Colloid Interface Sci.}, 2004,
  \textbf{270}, 113--122\relax
\mciteBstWouldAddEndPuncttrue
\mciteSetBstMidEndSepPunct{\mcitedefaultmidpunct}
{\mcitedefaultendpunct}{\mcitedefaultseppunct}\relax
\EndOfBibitem
\bibitem[Sholl \emph{et~al.}(2000)Sholl, Fenwick, Atman, and
  Prieve]{prieve2000shollSimulation}
D.~S. Sholl, M.~K. Fenwick, E.~Atman and D.~C. Prieve, \emph{J. Chem. Phys.},
  2000, \textbf{113}, 9268--9278\relax
\mciteBstWouldAddEndPuncttrue
\mciteSetBstMidEndSepPunct{\mcitedefaultmidpunct}
{\mcitedefaultendpunct}{\mcitedefaultseppunct}\relax
\EndOfBibitem
\bibitem[Nayeri \emph{et~al.}(2013)Nayeri, Abbas, and
  Bergenholtz]{noisevdwnayeri2013}
M.~Nayeri, Z.~Abbas and J.~Bergenholtz, \emph{Colloids Surf. A}, 2013,
  \textbf{429}, 74--81\relax
\mciteBstWouldAddEndPuncttrue
\mciteSetBstMidEndSepPunct{\mcitedefaultmidpunct}
{\mcitedefaultendpunct}{\mcitedefaultseppunct}\relax
\EndOfBibitem
\bibitem[Bitter \emph{et~al.}(2013)Bitter, Duncan, Beltran-Villegas,
  Fairbrother, and Bevan]{noisesoftenstrongforce2013Bevan}
J.~L. Bitter, G.~A. Duncan, D.~J. Beltran-Villegas, D.~H. Fairbrother and M.~A.
  Bevan, \emph{Langmuir}, 2013, \textbf{29}, 8835--8844\relax
\mciteBstWouldAddEndPuncttrue
\mciteSetBstMidEndSepPunct{\mcitedefaultmidpunct}
{\mcitedefaultendpunct}{\mcitedefaultseppunct}\relax
\EndOfBibitem
\bibitem[Mandel(1959)]{mandel1959poissonlight}
L.~Mandel, \emph{Proc. Phys. Soc. London}, 1959, \textbf{74}, 233--243\relax
\mciteBstWouldAddEndPuncttrue
\mciteSetBstMidEndSepPunct{\mcitedefaultmidpunct}
{\mcitedefaultendpunct}{\mcitedefaultseppunct}\relax
\EndOfBibitem
\bibitem[Ermak and McCammon(1978)]{displacementequation}
D.~L. Ermak and J.~A. McCammon, \emph{J. Chem. Phys.}, 1978, \textbf{69},
  1352--1360\relax
\mciteBstWouldAddEndPuncttrue
\mciteSetBstMidEndSepPunct{\mcitedefaultmidpunct}
{\mcitedefaultendpunct}{\mcitedefaultseppunct}\relax
\EndOfBibitem
\bibitem[Brenner(1961)]{brenner_nearwalldiffusion_verticle}
H.~Brenner, \emph{Chem. Eng. Sci.}, 1961, \textbf{16}, 242--251\relax
\mciteBstWouldAddEndPuncttrue
\mciteSetBstMidEndSepPunct{\mcitedefaultmidpunct}
{\mcitedefaultendpunct}{\mcitedefaultseppunct}\relax
\EndOfBibitem
\bibitem[Bevan and Prieve(2000)]{Prieve_hinderedDiffusion_2000}
M.~A. Bevan and D.~C. Prieve, \emph{J. Chem. Phys.}, 2000, \textbf{113},
  1228--1236\relax
\mciteBstWouldAddEndPuncttrue
\mciteSetBstMidEndSepPunct{\mcitedefaultmidpunct}
{\mcitedefaultendpunct}{\mcitedefaultseppunct}\relax
\EndOfBibitem
\bibitem[Verwey \emph{et~al.}(1948)Verwey, Overbeek, and
  Van~Nes]{verwey_doublelayertheory}
E.~J.~W. Verwey, J.~T.~G. Overbeek and K.~Van~Nes, \emph{Theory of the
  stability of lyophobic colloids: the interaction of sol particles having an
  electric double layer}, Elsevier Publishing Company, 1948\relax
\mciteBstWouldAddEndPuncttrue
\mciteSetBstMidEndSepPunct{\mcitedefaultmidpunct}
{\mcitedefaultendpunct}{\mcitedefaultseppunct}\relax
\EndOfBibitem
\bibitem[von Gr{\"u}nberg \emph{et~al.}(2001)von Gr{\"u}nberg, Helden,
  Leiderer, and Bechinger]{Bechinger_doublelayer}
H.-H. von Gr{\"u}nberg, L.~Helden, P.~Leiderer and C.~Bechinger, \emph{J. Chem.
  Phys.}, 2001, \textbf{114}, 10094--10104\relax
\mciteBstWouldAddEndPuncttrue
\mciteSetBstMidEndSepPunct{\mcitedefaultmidpunct}
{\mcitedefaultendpunct}{\mcitedefaultseppunct}\relax
\EndOfBibitem
\bibitem[Rudhardt \emph{et~al.}(1998)Rudhardt, Bechinger, and
  Leiderer]{rudhardt1998direct}
D.~Rudhardt, C.~Bechinger and P.~Leiderer, \emph{Phys. Rev. Lett.}, 1998,
  \textbf{81}, 1330\relax
\mciteBstWouldAddEndPuncttrue
\mciteSetBstMidEndSepPunct{\mcitedefaultmidpunct}
{\mcitedefaultendpunct}{\mcitedefaultseppunct}\relax
\EndOfBibitem
\bibitem[Leunissen \emph{et~al.}(2005)Leunissen, Christova, Hynninen, Royall,
  Campbell, Imhof, Dijkstra, van Roij, and van
  Blaaderen]{leunissen2005vanBlaadIonic}
M.~E. Leunissen, C.~G. Christova, A.-P. Hynninen, C.~P. Royall, A.~I. Campbell,
  A.~Imhof, M.~Dijkstra, R.~van Roij and A.~van Blaaderen, \emph{Nature}, 2005,
  \textbf{437}, 235--240\relax
\mciteBstWouldAddEndPuncttrue
\mciteSetBstMidEndSepPunct{\mcitedefaultmidpunct}
{\mcitedefaultendpunct}{\mcitedefaultseppunct}\relax
\EndOfBibitem
\bibitem[Hueckel \emph{et~al.}(2020)Hueckel, Hocky, Palacci, and
  Sacanna]{sacanna2020ionicTed}
T.~Hueckel, G.~M. Hocky, J.~Palacci and S.~Sacanna, \emph{Nature}, 2020,
  \textbf{580}, 487--490\relax
\mciteBstWouldAddEndPuncttrue
\mciteSetBstMidEndSepPunct{\mcitedefaultmidpunct}
{\mcitedefaultendpunct}{\mcitedefaultseppunct}\relax
\EndOfBibitem
\bibitem[Hertlein \emph{et~al.}(2008)Hertlein, Helden, Gambassi, Dietrich, and
  Bechinger]{bechinger2008casimirTIRM}
C.~Hertlein, L.~Helden, A.~Gambassi, S.~Dietrich and C.~Bechinger,
  \emph{Nature}, 2008, \textbf{451}, 172--175\relax
\mciteBstWouldAddEndPuncttrue
\mciteSetBstMidEndSepPunct{\mcitedefaultmidpunct}
{\mcitedefaultendpunct}{\mcitedefaultseppunct}\relax
\EndOfBibitem
\bibitem[Rogers and Crocker(2011)]{RogersPNAS}
W.~B. Rogers and J.~C. Crocker, \emph{Proc. Natl. Acad. Sci. U.S.A.}, 2011,
  \textbf{108}, 15687--15692\relax
\mciteBstWouldAddEndPuncttrue
\mciteSetBstMidEndSepPunct{\mcitedefaultmidpunct}
{\mcitedefaultendpunct}{\mcitedefaultseppunct}\relax
\EndOfBibitem
\bibitem[Doye \emph{et~al.}(1995)Doye, Wales, and Berry]{doye1995effect}
J.~P. Doye, D.~J. Wales and R.~S. Berry, \emph{J. Chem. Phys.}, 1995,
  \textbf{103}, 4234--4249\relax
\mciteBstWouldAddEndPuncttrue
\mciteSetBstMidEndSepPunct{\mcitedefaultmidpunct}
{\mcitedefaultendpunct}{\mcitedefaultseppunct}\relax
\EndOfBibitem
\bibitem[Bou-Rabee and Holmes-Cerfon(2020)]{bou2020sticky}
N.~Bou-Rabee and M.~C. Holmes-Cerfon, \emph{SIAM Review}, 2020, \textbf{62},
  164--195\relax
\mciteBstWouldAddEndPuncttrue
\mciteSetBstMidEndSepPunct{\mcitedefaultmidpunct}
{\mcitedefaultendpunct}{\mcitedefaultseppunct}\relax
\EndOfBibitem
\bibitem[Eichmann \emph{et~al.}(2013)Eichmann, Meric, Swavola, and
  Bevan]{protein--carbohydrateBevan2013}
S.~L. Eichmann, G.~Meric, J.~C. Swavola and M.~A. Bevan, \emph{Langmuir}, 2013,
  \textbf{29}, 2299--2310\relax
\mciteBstWouldAddEndPuncttrue
\mciteSetBstMidEndSepPunct{\mcitedefaultmidpunct}
{\mcitedefaultendpunct}{\mcitedefaultseppunct}\relax
\EndOfBibitem
\end{mcitethebibliography}
    \bibliographystyle{rsc} %the RSC's .bst file
\end{document}